\documentclass{PoS}
\usepackage[authoryear,square]{natbib}
\bibpunct{(}{)}{;}{a}{}{,}

\usepackage{amsmath}

\newcommand{\Hu}{{\cal H}}

\title{Measuring redshift-space distortions with future SKA surveys}

\ShortTitle{RSD with the SKA}

\author{Alvise Raccanelli$^{1,2,3}$, Philip Bull$^{4}$, Stefano Camera$^{5,6}$, David Bacon$^{7}$, Chris Blake$^{8}$, Olivier Dor\'{e}$^{2,3}$, Pedro Ferreira$^{9}$, Roy Maartens$^{10,7}$, Mario Santos$^{10,11,6}$, Matteo Viel$^{12,13}$, Gong-bo Zhao$^{14,7}$
\\
$^1$ Department of Physics \& Astronomy, Johns Hopkins University, 3400 N. Charles St., Baltimore, MD 21218, USA \\
$^2$ Jet Propulsion Laboratory, California Institute of Technology, Pasadena CA 91109, USA \\
$^3$ California Institute of Technology, Pasadena CA 91125, USA \\
$^4$ Institute of Theoretical Astrophysics, University of Oslo, P.O. Box 1029 Blindern, N-0315 Oslo, Norway \\
$^5$ Jodrell Bank Centre for Astrophysics, The University of Manchester, Manchester M13 9PL, UK
$^6$ CENTRA, Instituto Superior T\'{e}cnico, Universidade de Lisboa, Lisboa 1049-001, Portugal \\
$^7$ Institute of Cosmology \& Gravitation, University of Portsmouth, Portsmouth PO1 3FX, UK \\
$^8$ Centre for Astrophysics \& Supercomputing, Swinburne University of Technology, PO Box 218, Hawthorn, VIC 3122, Australia \\
$^{9}$ Astrophysics, University of Oxford, DWB, Keble Road, Oxford OX1 3RH, UK \\
$^{10}$ Physics Department, University of the Western Cape, Cape Town 7535, South Africa \\
$^{11}$ SKA SA, 4rd Floor, The Park, Park Road, Pinelands, 7405, South Africa \\
$^{12}$ INAF - Osservatorio Astronomico di Trieste, Via Tiepolo 11, 34143, Trieste, Italy \\
$^{13}$ INFN sez. Trieste, Via Valerio 2, 34127 Trieste, Italy \\
$^{14}$ National Astronomy Observatories, Chinese Academy of Science, Beijing, 100012, P.R.China \\ \\
E-mail: \email{alvise@jhu.edu}
}

\abstract{
The peculiar motion of galaxies can be a particularly sensitive probe of gravitational collapse. As such, it can be used to measure the dynamics of dark matter and dark energy as well the nature of the gravitational laws at play on cosmological scales. Peculiar motions manifest themselves as an overall anisotropy in the measured clustering signal as a function of the angle to the line-of-sight, known as redshift-space distortion (RSD). Limiting factors in this measurement include our ability to model non-linear galaxy motions on small scales and the complexities of galaxy bias. The anisotropy in the measured clustering pattern in redshift-space is also driven by the unknown distance factors at the redshift in question, the Alcock-Paczynski distortion. This weakens growth rate measurements, but permits an extra geometric probe of the Hubble expansion rate.
In this chapter we will briefly describe the scientific background to the RSD technique, and forecast the potential of the SKA phase 1 and the SKA2 to measure the growth rate using both galaxy catalogues and intensity mapping, assessing their competitiveness with current and future optical galaxy surveys. }

\FullConference{
Advancing Astrophysics with the Square Kilometre Array\\
June 8-13, 2014\\
Giardini Naxos, Italy}

\begin{document}

%%%%%%%%%%%%%%%%%%%%%%%%%%%%%%%%%%%%%%%%%%%%%%%%%%%%%%%%%%%%%%%%%%%%%%%%%%%%%%%%%%%%%%%%%%%%%%%%%%%%%%%%%%%%%%%%%%%%%%%%%%%%%%%%%%%%%%%%%%%%%%%%%%%			INTRODUCTION			%%%%%%%%%%%%%%%%%%%
%%%%%%%%%%%%%%%%%%%%%%%%%%%%%%%%%%%%%%%%%%%%%%%%%%%%%%%%%%%%%%%%%%%%%%%%%%%%%%%%%%%%%%%%%%%%%%%%%%%%%%%%%%%%%%%%%%%%%%%%%%%%%%

\section{Introduction}
One of the biggest challenges of modern cosmology is to understand the accelerated expansion rate of the Universe. A number of proposals have been put forward, the most notable of which are the presence of dark energy or, alternatively, a modification to general relativity on cosmological scales. 
Radio surveys have been used to test these different hypotheses in the past, mainly using the integrated Sachs-Wolfe effect and the galaxy angular power spectrum~\citep[see e.g.][]{Nolta:2004, Raccanelli:2008nx, Xia:2010}. With the new generation of radio arrays such as LOFAR~\citep{lofar} and ASKAP~\citep{askap}, and in preparation to the SKA, there is a growing interest in understanding how future radio surveys can constrain cosmological parameters, and maybe discriminate between the two scenarios described above. Some examples of these investigations can be found in~\cite{Raccanelliradio, Camera:2012ly, Raccanelli:2014isw}.
Here we focus on how the proposed SKA surveys will be able to provide measurements of Redshift-Space Distortions (RSD), allowing in this way measurements of cosmological parameters, and in particular the growth of structures.

In this chapter, we will briefly review the physics of Redshift-Space Distortions and how we model their effect on the power spectrum. We investigate potential issues and systematics and then we present forecasts for the measurements we can perform using the different proposed SKA surveys, focusing in particular on models describing the growth of structures.
In two complementary chapters~\citep{Bull, Camera} are presented forecasts for the BAO and large-scale measurements, investigating how those can constrain dark energy models and primordial non-Gaussianity.

\section{The physics of the growth rate}
The presence of a dark energy component in the energy-density of the Universe (or the fact that our theory of gravity needs to be modified on large scales), modifies the gravitational growth of large-scale structures. The large-scale structure we see traced by the distribution of galaxies arises through gravitational instability, which amplifies primordial fluctuations that originated in the very early Universe.
The rate at which structure grows from small perturbations offers a key discriminant between cosmological models, as different models predict measurable differences in the growth rate of large-scale structure with cosmic time~\citep[e.g.][]{Jain:2007, Song:2008a, Song:2008b}. For instance, dark energy models in which general relativity is unmodified predict different large-scale structure formation compared to Modified Gravity models with the same background expansion~\citep[e.g.][]{Dvali:2000, Carroll:2004, Brans:2000, Yamamoto:2008, Yamamoto:2010}.

The growth rate, $f(a)$, as a function of scale factor $a$ is defined as:
\begin{eqnarray}
\label{eq:fdef}
f(a)&\equiv&\frac{d\ln \delta_M(a)}{d\ln a} \, ,
\end{eqnarray}
where $\delta_M(a)$ is the amplitude of the growing mode of matter density perturbations. In the conformal Newtonian gauge the evolution equations for the velocity potential $\theta$ and $\delta_M$ are:
\begin{eqnarray}
\dot{\delta}_M&=&3(\dot\Phi+\Hu\Psi)-\left[k^2+3(\Hu^2-\dot\Hu)\right]\theta_M \, , \label{Delta_evol}\\
\dot{\theta}_M&=&-\Hu\theta_M+\Psi \, . \label{theta_evol}
\end{eqnarray}
We use dots to denote derivatives with respect to conformal time, and our conventions for the metric potentials are displayed in the perturbed line element: $
ds^2=a^2(\eta)\left[-(1+2\Psi)d\eta^2+ (1-2\Phi)dx^idx_i \right]$ which satisfy (in the quasi-static regime) the field equations: $
2\nabla^2\Phi=\kappa a^2\,\mu(a,k)\,{\bar \rho}_M\Delta_M$ and ${\Phi}/{\Psi}=\gamma(a,k)$. The parameters $\mu$ and $\gamma$ encapsulate all the information about deviation from GR for metric theories of gravity~\citep{Baker:2011}.

Again, in the quasi static regime, the evolution equation for $\delta_M$ simplifies to:
\begin{eqnarray} 
\ddot\delta_M+\Hu\dot{\delta}_M-\frac{3}{2}{\cal H}^2\Omega_M\xi{\delta}_M=0 \, , \label{DeltaF}
\end {eqnarray}
where we have defined $\xi\equiv\mu/\gamma$ (which is equal to $1$ in GR). Using $x=\ln a$ as the independent variable we have:
\begin{eqnarray}
\delta^{''}_M+\left(1+\frac{\Hu^{'}}{\Hu}\right){\delta}^{'}_M-\frac{3}{2}\Omega_M\xi{\delta}_M=0 \, . \label{DeltainF}
\end {eqnarray}
Primes denote derivatives with respect to $x$. We can convert this into an evolution equation for $f$ :
\begin{eqnarray}
& f^\prime+q(x)\,f+f^2=\frac{3}{2}\Omega_M \xi \label{f_eq} \, , \label{qdef}
\end{eqnarray}
where $q(x)=\frac{1}{2}[1-3\,\omega(x) (1-\Omega_M(x))]$. The effect of the expansion rate (via $q(x)$) and modified gravity (via $\xi$) are explicit in the time evolution of $f$.

\section{Redshift Space Distortions}
\label{sec:rsd}
Measurements of RSD played an important role in developing the current cosmological model, and it will be a fundamental part of several future cosmological experiments, because observations of RSD in galaxy surveys are a powerful way to study the pattern and the evolution of the Large Scale Structure of the Universe \citep{Kaiser:1987, Hamilton:1997}, as they provide constraints on the amplitude of peculiar velocities induced by structure growth, thereby allowing tests of the theory of gravity governing the growth of those perturbations.
RSD have been measured using techniques based on both correlation functions and power-spectra \citep[e.g.][]{Peacock:2001, Percival:2004, Tegmark:2006, Guzzo:2008, Samushia:2011wa, Samushia:2013, Reid:2010, Reid:2012, Sanchez:2012, Blake:2010, Blake:2011, Blake:2012}; the most recent analyses come from BOSS DR11 and GAMA~\citep{Samushia:2014, Blake:2013}.

\subsection{Formalism}
\label{sec:rsd}
RSD arise because we infer galaxy distances from their redshifts using the Hubble law: the radial component of the peculiar velocity of individual galaxies will contribute to each redshift and will be misinterpreted as being cosmological in origin, thus altering our estimate of the distances to them.
The correction due to peculiar velocities can be used to set constraints on cosmological models and parameters, as it depends on the coherent large scale infall of matter toward overdense regions.
The relation between the redshift-space position $\mathbf{s}$ and real-space position $\mathbf{r}$  is:
\begin{equation}
\label{eq:stor}
\mathbf{s}({\bf{r}} ) = {\bf{r}} + v_r ({\bf{r}} ) \hat {\bf{r}},
\end{equation}
where $v_r$ is the velocity in the radial direction.

The Redshift-Space Distortions (RSD) corrections come from the fact that the real-space position of a source in the radial direction in modified by peculiar velocities due to local overdensities; this effect can be modeled as~\citep{Kaiser:1987, Hamilton:1997}:
\begin{equation}
\label{eq:kaiser}
\delta^s(k) = \left( 1+\beta \mu^2 \right) \delta^r(k) \, ,
\end{equation}
where, in the linear regime, $\beta$ is the quantity that solves the linearized continuity equation:
\begin{equation}
\beta\delta + \bar{\nabla} \cdot \bar{v} = 0 \, .
\end{equation}
Here $\beta = f/b$, where $b$ is the bias relating the visible to the underlying matter distribution (see Section~\ref{sec:bias} for more details on it).
%, and the parameter $f$ is defined as the logarithmic derivative of the growth factor with respect to the scale factor $a$:
%\begin{equation}
%\label{eq:f}
%f = \frac{d \,\,  \rm ln \, D}{d \,\,  \rm ln \, a} \, ,
%\end{equation}
%where $D(a) \propto \delta_m$ ($\delta_m$ being the fractional matter density perturbation).
%%

For this reason, measuring $f$ from RSD allows us to set constraints on cosmological models and parameters.

%\subsection{Constraining $f$ and $f\sigma_8$}
\subsection{The power spectrum}
\label{sec:pow}
The matter power spectrum depends on a variety of cosmological parameters, and for this reason its measurement has been used (together with its Fourier transform, the correlation function) to constrain e.g. dark energy parameters~\citep{Samushia:2011wa}, models of gravity~\citep{Raccanelligrowth}, neutrino mass~\citep{dePutter:2012, Zhao:2012}, dark matter models~\citep{Cyr-Racine:2014, Dvorkin:2014}, the growth of structures~\citep{Samushia:2013, Reid:2012}, and non-Gaussianity~\citep{Ross:2013}. 

We define the power spectrum as:
\begin{align}
\label{eq:kaiser}
P^s_g(k,\mu,z) = \left[ b (z) + f(z) \mu^2 \right]^2 P_m^r(k,z) + P_{shot}(z) \, ,
\end{align}
where the superscripts $^r$ and $^s$ indicate real and redshift-space, respectively, and the subscripts $_m$ and $_g$ stands for matter and galaxies; $\mu$ is the angle with the line of sight. The shot noise contribution is taken to be:
\begin{align}
P_{shot}(z) = \frac{1}{\bar{n}_g(z)} \, .
\end{align}

%In the next sub-sections we will discuss a number of corrections that should be made to the power spectrum modeling when doing a careful real data analysis and list some of the possible systematics that can potentially affect the reliability of such measurements.

The standard analysis of RSD makes use of the so-called Kaiser formalism (Equation~\ref{eq:kaiser}), that relies on several simplifying assumptions, including considering only the linear regime and the distant observer approximation; in Section~\ref{sec:beyond} we briefly mention some possible extensions of this model. In this Chapter we will make use of the Kaiser formula, but for a detailed data analysis some further investigations will be needed.

%, so limiting the range of scales of applicability of this modeling. There have been several attempts to model smaller scales RSD, exploring the quasi-linear regime~\citep[e.g.][]{Scoccimarro:2004, Taruya:2009, Taruya:2010, Reid:2010, Kwan:2011}; on large scales, recently~\cite{Bertacca:2012} developed a formalism to compute the correlation function including GR corrections, that arise when probing scales comparable to the Hubble scales (see Section~\ref{sec:large} for more details and references). In this paper we start from the standard Kaiser formalism, and we briefly explore some extensions of the power spectrum modeling.

\subsection{Bias}
\label{sec:bias}
While the distribution of galaxies is the observed quantity, the cosmological model directly predicts the statistical distribution of (dark) matter. The simplest assumption is that the galaxy
distribution is a biased version of the underlying matter field, the
so-called linear bias model, at position {\bf{x}}:
\begin{equation}
\label{eq:biaslin}
\delta_{\rm galaxies}({\bf{x}})= b\, \delta_{\rm matter}({\bf{x}})\,, 
\end{equation}
with $b$ a constant bias factor independent of a given
smoothing scale $R$ over which the density fields are calculated.
This model is motivated by the fact that rare peaks in the density
field (e.g. clusters of galaxies) have to be more strongly
clustered (i.e. biased) than matter itself ~\citep[e.g.][]{Kaiser:1984}. 
This is a simplifying assumption and the
relationship between galaxies and matter is more complex: in fact,
clustering properties of galaxies do depend on galaxies' intrinsic
features. For example the relation could be scale-dependent,
nonlinear, stochastic, non-local, a function of the particular sample of galaxies chosen, a
function of cosmic time or dependent on many other physical quantities
(such as the gas temperature, environment, merging history, etc.).
Thus, the above equation can be generalized to a more complex form:
\begin{equation}
\delta_{\rm galaxies}({\bf{x}})= f(\delta_{\rm matter}({\bf{x}})+\epsilon) \, ,
\end{equation}
with $\epsilon$ embedding all the dependencies on physical quantities
other than dark matter density.

Currently, analytical efforts to model the bias are first attempting to model the
halo-matter bias by relying on: the peak background split formalism in
a coarse grained perturbation theory framework~\citep[e.g.][and references therein]{Schmidt:2013};
the excursion set approach also for
non gaussian initial conditions as in~\cite{Musso:2012};
perturbation theories~\citep{Bernardeau:2002}.  Particular emphasis
is also put on unveiling the scale dependence of the bias, out to the
largest scales, that can be a powerful probe for testing initial
conditions and/or the nature of gravity and it has been recently shown
that also in the standard cosmological model ($\Lambda$CDM) the halo
bias is scale dependent due to general relativitistic effects and
not only to non-gaussianities~\citep[e.g.][]{Baldauf:2011}.

%In the
%linear regime the evolution at time $t$ of the bias of haloes of mass
%M identified at time $t_1$ can be easily described by the following:
%\begin{equation}
%b_h(M_1,\delta_1,;t)=1+\frac{1}{D(t)}\left(\frac{\nu_1^2-1}{\delta_1}\right)\,,
%\end{equation}
%with $\nu_1=\delta_1/\sqrt{S_1}$, $S_1=\sigma^2(M)$ the mass variance
%of the smoothed density field that together with the growth
%factor $D(t)$ entails the cosmology dependence.

%Some scale dependence between matter and halo bias can also be induced
%by massive neutrino, at intermediate scales that are mildly affected
%by non-linearities (Villaescusa-Navarro et al. 2014), that could
%impact also on the redshift space distortion signal (Marulli et
%al. 2011).
A comprehensive analysis of halo bias is presented in~\cite{Smith:2006} by comparing the results of N-body simulations
with semi-analytical prescriptions based on perturbation theory and
the halo model, also relying on the cross-spectrum between matter and
haloes. In the work above, it is shown that the non-linearities of the
bias are determined not only by the non-linear evolution of the power
spectrum but also by the fact that haloes of different masses are
biased in a different way. 
In a recent study based on N-body
simulations complemented by a galaxy formation model,~\cite{Crocce:2013}
found a nearly scale independent bias at the level of $\sim
2-5$ \% at scales larger than 20 Mpc$/h$ for a mock Luminous Red
Galaxies sample.

%These recent works highlight the fact that modeling of the
%galaxy-matter relation is usually performed at a second stage
%(post-processing of N-body simulations) either with halo occupation
%distribution modeling or the so-called sub-halo abundance matching
%techniques that assign galaxies to dark matter haloes in order to
%reproduce some observed statistical properties of the galaxies
%themselves. This step encodes the details of the galaxy formation
%model (for a recent review see~\cite{Baugh:2013}.

Overall, simulations show that on scales larger than $\approx 30$ Mpc/h the simplest linear parameterization works reasonably well,
so for the purposes of this Chapter, we will assume the bias to be linear
and constant at the scales of interests for SKA forecasts as in Equation~\ref{eq:biaslin}; however, this assumption should be carefully tested in the future.

\subsection{Beyond the Kaiser model}
\label{sec:beyond}
Equation~\ref{eq:kaiser} is valid only on linear scales, assumes the plane-parallel approximation and is derived using Newtonian physics; this approximation is valid when considering pair separations in a limited range of scales, large enough to avoid non-linear effect (i.e. $\gtrsim 30$ Mpc) and are relatively small (up to $\sim 200$ Mpc). If one wants to extend analyses of RSD to smaller and larger scales, there are modifications to the standard formalism to take into account.

\subsubsection{Non-linearities}
\label{sec:nl}
Within dark matter haloes, peculiar velocities of galaxies are highly non-linear, and these velocities can induce RSD that are larger than the real-space distance between galaxies within the halo. For this reason, on small scales we observe the so-called Fingers of God (FOG) effect -- strong elongation of structures along the line of sight~\citep{Jackson:1972}.
This results in a damping of the power spectrum on small scales compared to the predictions of the linear model, and is usually modeled by multiplying the linear power-spectrum by a function $F(\sigma_{ v},k,\mu)$, where $\sigma_{v}$ is the average velocity dispersion of galaxies within the relevant haloes. 

Modeling of non-linearities has been investigated numerous times~\citep[e.g.][]{Scoccimarro:2004, Taruya:2009, Taruya:2010, Reid:2010, Anselmi:2010, Anselmi:2012, Kwan:2011, Neyrinck:2009, Neyrinck:2011, Jennings:2012, Carron:2013}.

%For the purpose of this paper we limit ourselves to the linear regime and look into the mildly quasi-linear regime and how we can gain in terms of cosmological constraints, using a damping of the power spectrum as in~\cite{Samushia:2011wa}, using the two most frequently used functions (see e.g.~\cite{Cole:1994, Peacock:1996}):
%\begin{eqnarray}
%  F_{\rm Lorentzian}(k,\mu^2) &=& \left[1+(k\sigma_{v}\mu)^2\right]^{-1} 
%    \label{eq:F_exp}, \\
%    F_{\rm Gaussian}(k,\mu^2) &=& \exp\left[-(k\sigma_{v}\mu)^2\right]
%    \label{eq:F_gauss}.
%\end{eqnarray}

In this Chapter we look only at linear scales; extensions to the quasi- and non- linear regimes will help giving more constraining power, but they require investigations that are beyond the scope of this paper.

\subsubsection{Large scale effects}
\label{sec:large}
When considering wide surveys and galaxy pairs with large separation, a more precise analysis involving wide-angle and GR corrections should be used~\citep[see e.g.][]{Szalay:1997, Matsubara:1999, Szapudi:2004, Papai:2008, Raccanelli:2010fk, Samushia:2011wa, Montanari:2012, Bertacca:2012, Raccanelligrowth, Raccanelli:2013}. 
Moreover, on very large scales, the modeling for the power spectrum needs to take into account General Relativity (GR) effects that will be important on scales comparable to the horizon~\citep[see e.g.][]{Yoo:2010, Bonvin:2011, Challinor:2011, Yoo:2012, Jeong:2011, Bertacca:2012, Raccanelliradial, DiDio:2014}.

However, including wide-angle and GR corrections in the power spectrum is beyond the scope of this Chapter.
A more detailed analysis of large scale effects for the SKA is carried out in the SKA Chapter ``Cosmology on the Largest Scales''~\citep{Camera}.

\subsection{Alcock-Paczynski Effect}
Positions of galaxies are given in terms of angular positions and redshifts; angular diameter distances and Hubble expansion rates as functions of redshift are required in order to convert angular and redshift separations into physical distances. Those functions depend on the adopted cosmological model.
If the real cosmology is significantly different from the fiducial one, this difference will introduce additional anisotropies in the correlation function through the Alcock-Paczynski effect. This can significantly bias the measurements of growth~\citep[see e.g.][]{Ballinger:1996, Simpson:2010, Samushia:2011wa, Montanari:2012}.

In the presence of Alcock-Paczynski effect, the redshift-space power-spectrum is:
\begin{eqnarray}
  P^{\rm s}(k',\mu',\alpha_\bot,\alpha_{||},{\bf p})=
  \frac{(b+\mu'^{2}f)^2} {\alpha_{\bot}^2\alpha_{||}} 
  P^{\rm r}\left[\frac{k'}{\alpha_\bot}\sqrt{1+\mu^{'2}\left(\frac{1}{F^2}-1\right)}\right],
  \label{eq:rsdap}
\end{eqnarray}
\noindent
where ${\bf p}$ are standard cosmological parameters determining the shape of
the real-space power-spectrum, $k'$ and $\mu'$ are the observed wavevector and
angle, related to the real quantities by
  $k'_{||}=\alpha_{||}k_{||}$,
  $k'_\bot=\alpha_\bot k_\bot$,
  $\mu'=\frac{k_{||}'}{\sqrt{k_{||}'+k_\bot'}}$,
where $F=\alpha_{||}/\alpha_\bot$, with $\alpha_{||}$ and $\alpha_\bot$ being the ratios of angular and radial
distances between fiducial and real cosmologies,
  $\alpha_{||}=\frac{H^{\rm fid}}{H^{\rm real}}$,
  $\alpha_\bot=\frac{D^{\rm real}}{D^{\rm fid}}$.

Ignoring the AP effect is equivalent to assuming that $\alpha$ factors
are equal to unity in Eq.~(\ref{eq:rsdap}).

\section{SKA Surveys}
\label{sec:surveys}

The Square Kilometre Array (SKA) project is an international effort to build the world's largest radio telescope, several times more sensitive than any existing radio telescope and capable of addressing fundamental questions about the Universe~\citep{Carilli:2004}. The SKA will be developed in two stages. The first stage currently encompasses two mid-frequency facilities ($\sim$ 1 GHz) operating within South Africa (SKA1-Mid) and Australia (SKA1-Sur). A low frequency array (SKA1-Low $\sim$ 100 MHz) will also be set in Australia. We refer to~\cite{Dewdney:2009} for a description of the setups. In the second stage of the SKA, the plan is
to extend the array by about a factor of 10, both in collecting area and primary beam (field of view), thus significantly increasing the survey power of the facility. In the following sections, we consider two types of surveys that can be used to probe the redshift space distortions.

\subsection{HI surveys}
\label{sec:HI}
The most straightforward way to go after the RSD signal is through a line galaxy survey. In the radio, the solution is to use the HI 21cm line which, by measuring its characteristic shape, will allow determination of very accurate redshifts ($\delta z < 1.0\times 10^{-4}$). The advantage of such threshold surveys is that we can be confident to be free of any foreground contamination. The disadvantage is that it requires high sensitivities to detect HI galaxies at non-local redshifts (the highest redshift HI galaxies detected up to date was at $z\sim 0.14$ with Arecibo~\citep{Arecibo}).

\begin{figure*}[htb!]
\includegraphics[width=0.49\linewidth]{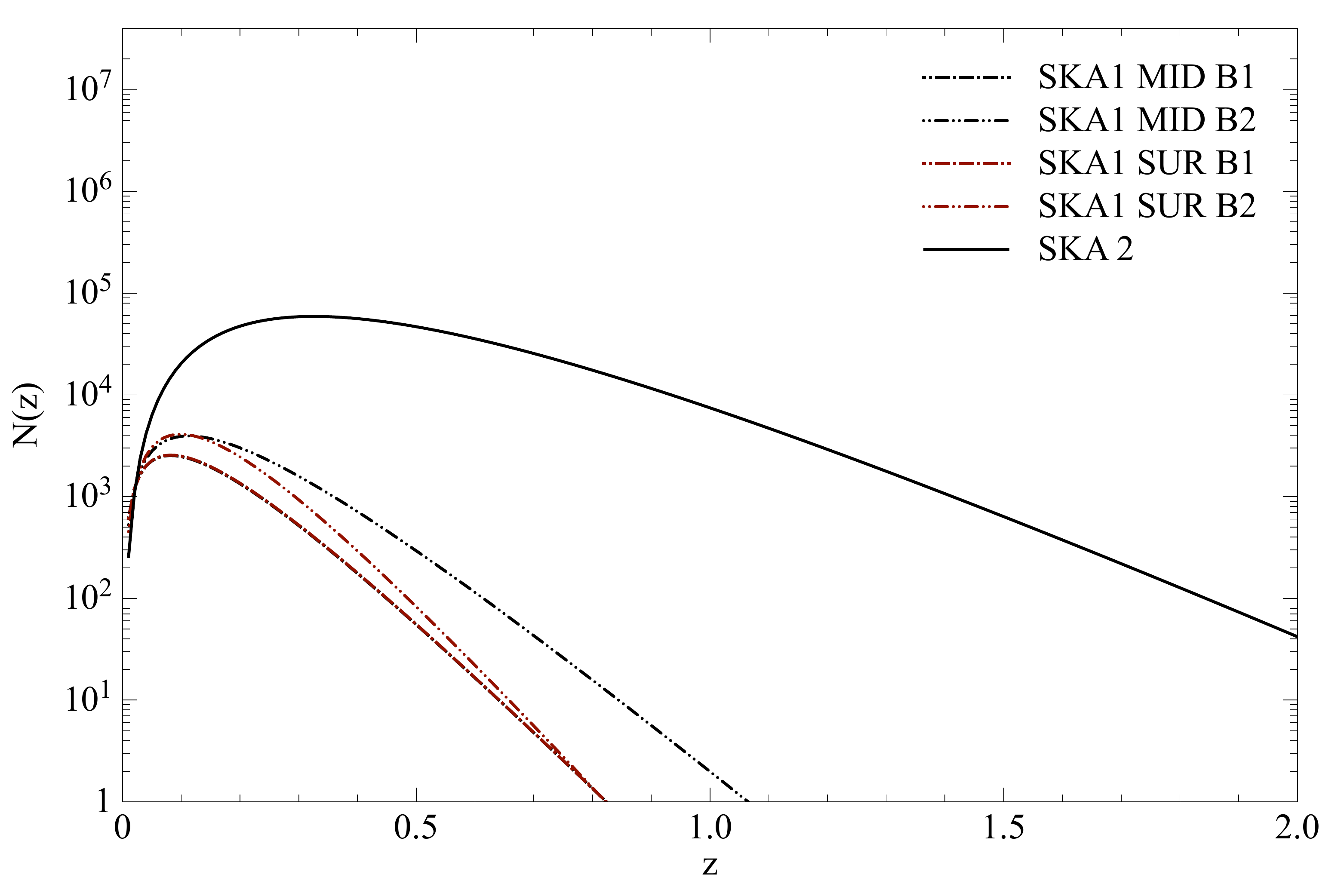}
\includegraphics[width=0.49\linewidth]{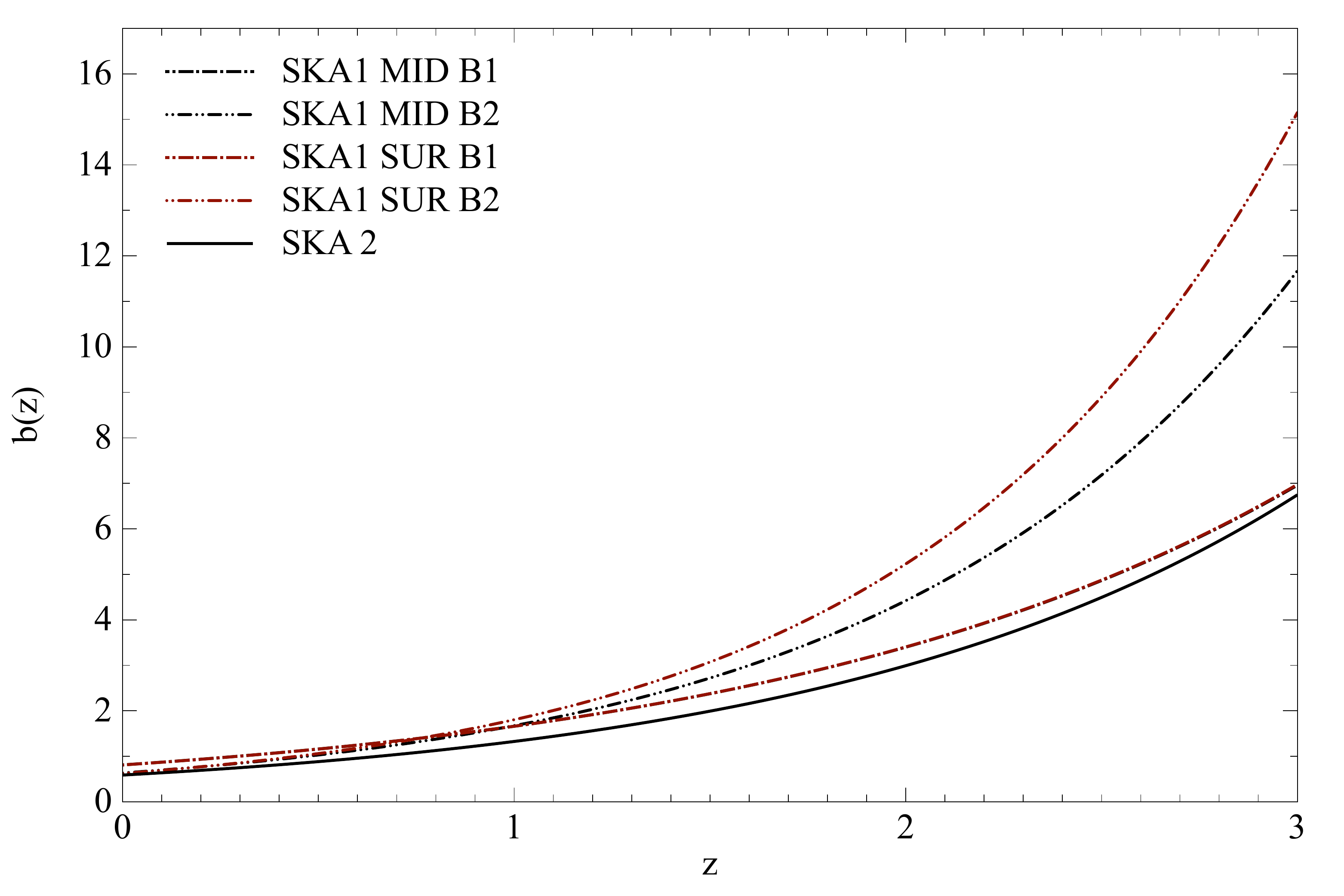}
\caption{Redshift distributions ({\it left panel}) and bias ({\it right panel}) for the SKA1 and SKA2 surveys used in this work.
}
\label{fig:Nz}
\end{figure*}

Cosmological applications will require detecting enough galaxies to beat shot noise and over a large enough area to reduce cosmic variance.  With the sensitivities for SKA1 and taking 10,000 hours of observation time, the optimal survey area will be around 5,000 deg$^2$. This will allow the detection of about $10^7$ galaxies using band 2 from SKA1-Mid or SKA1-Sur, while SKA1-Mid band 1 should detect less galaxies ($\sim 10^4$) since it will be constrained to higher redshifts ($0.4 \lesssim z \lesssim 3$). SKA2, on the other hand, should be capable of detecting about $10^9$ galaxies over a 30,000 deg$^2$ area, up  to $z\sim 2.0$, making it the largest galaxy redshift survey ever. The noise calculations and parameters for this HI galaxy survey can be found in the HI simulations chapter~\citep{SKA:HIsims}.
In Figure~\ref{fig:Nz} we plot the redshift distributions and bias for the different SKA configurations described above.
\subsection{Late-time HI intensity mapping} 

A relatively new alternative to large galaxy redshift surveys is {21 cm \it intensity mapping}. Galaxy surveys need high signal-to-noise detections of many millions of individual sources, requiring high flux sensitivity and long, dedicated surveys to reach $z \sim 1$. Intensity mapping (IM) attempts to circumvent these requirements by performing fast, low angular resolution surveys of the redshifted 21 cm emission line from neutral hydrogen (HI) integrated over many unresolved galaxies. For a more extensive discussion on HI intensity mapping, particularly in the context of the SKA, we refer to \citet{Santos:HIIM}. If we assume that, after reionisation, all the neutral hydrogen is contained within galaxies, as host galaxies are biased tracers of the cosmological large scale structure, so too is the integrated HI emission. Much of the cosmological information of interest (e.g. RSDs and BAOs) is found at large scales, so the lack of resolution is tolerable, and as the signal is from an emission line, redshift information is automatically provided as well. This allows large surveys to be performed extremely rapidly, efficiently recovering the 3D redshift-space matter power spectrum on large scales. An intensity mapping survey on SKA1-MID or SUR will be able to measure BAOs and RSDs over 25,000 deg$^2$ on the sky from $0 \lesssim z \lesssim 2.5$, for example \citep{Santos:HIIM}.

One way of thinking about an IM survey, then, is as a galaxy survey with the small angular scales averaged out. Information in the radial direction is mostly preserved, as modern radio receivers have sufficiently narrow frequency channel bandwidths that high redshift resolution can be obtained. The model for the RSD signal in intensity maps is therefore quite similar to that for a galaxy survey, except that the observable is the power spectrum of HI brightness temperature fluctuations, $\langle \delta T^*_b \delta T_b \rangle \propto T^2_b P(\mathbf{k})$, where $T_b$ is the mean HI brightness temperature. Note that the shot noise contribution has to be replaced by a more complicated direction-dependent noise term \citep[see e.g.][]{Bull:2014rha}. In this chapter we will focus on the RSD constraints that can be achieved by measuring the anisotropic power spectrum with IM surveys on SKA1-MID and SUR. Our forecasts are for 10,000 hour autocorrelation surveys over 25,000 deg$^2$ on bands 1 and 2 of both arrays.

\section{Forecast}
\label{sec:forecast}
In this Section we forecast the cosmological measurements that will be performed using the SKA using the configuration presented in Section~\ref{sec:surveys}; we present forecasts on parameters describing models for the growth of structures mentioned in Section~\ref{sec:growth}.

\subsection{Fisher Analysis}
\label{sec:fisher}
In order to predict the precision in the measurements of cosmological parameters, we perform a Fisher matrix analysis~\citep{Fisher:1935, Tegmark:1997};
we write the curvature or Fisher matrix for the power spectrum in the following way:
\begin{align}
\label{eq:FM}
F_{\alpha\beta} = \int_{z_{\rm min}}^{z_{\rm max}} dz \int_{k_{\rm min}}^{k_{\rm max}}dk  \int_{-1}^{+1}d\mu
& \left[\frac{\bar{n}_g(z) P(k,\mu,z)}{1+\bar{n}_g(z) P(k,\mu,z)}\right]^2
\frac{V_s(z) k^2}{8\pi^2 \left[P(k,\mu,z)\right]^2} \frac{\partial P(k,\mu,z)}{\partial \vartheta_\alpha}\frac{\partial P(k,\mu,z)}{\partial \vartheta_\beta} B_{nl} \, ,
\end{align}
where $\vartheta_{\alpha(\beta)}$ is the $\alpha(\beta)$-th cosmological parameter, $V_s$ is the volume of the survey and $\bar{n}_g$ is the mean comoving number density of galaxies.
The last term accounts for the non-linearities induced by the BAO peak~\citep{Seo:2006}:
\begin{equation}
B_{nl} = e^{-k^2\Sigma_{\perp}^2 -k^2 \mu^2 \left( \Sigma_{||}^2 -\Sigma_{\perp}^2 \right) } ,
\end{equation}
and $\Sigma_\bot=\Sigma_0D$, $\Sigma_{||}=\Sigma_0(1+f)D$, where $\Sigma_0$ is a constant
phenomenologically describing the nonlinear diffusion of the BAO peak due to
nonlinear evolution. From N-body simulations its numerical value is 12.4 $\rm h^{-1} Mpc$ and seems to depend linearly on $\sigma_8$, but only weakly on $k$ and cosmological parameters.
The integral in $k$ is performed in each redshift bin using~\citep{Smith:2003}:
\begin{align}
k_{\rm min} &= \frac{2 \pi}{V_{\rm bin}^{1/3}} \, ;\\
k_{\rm max} &= k_{\rm \, NL,0} (1+z)^{2/(2 + n_s)} \, .
\end{align}

In the rest of this Section we present the models we investigate: we focus on ways to explain the cosmic acceleration, either via dynamical dark energy or modifications of the model for gravity.

\subsection{Growth of structures}
\label{sec:growth}
We study how the SKA could constrain parameters describing the growth of structures. There are several models for it based on different explanations for the accelerated expansion of the Universe; they can be divided into two main categories: dark energy and modified growth.
Measuring RSD allows us to test different cosmological models and provides a good discriminant between modified gravity and dark energy models~\citep[see e.g.][]{Linder:2005, Linder:2007, Guzzo:2008}.

%deviations from General Relativity via measurements of the parameter $\gamma$.
%In the left panel of Figure~\ref{fig:gamma} we show the expected constraints on the growth rate.
%% (or $f \sigma_8$) [...]

\subsubsection{Dark Energy Models}
\label{sec:de}
In the standard $\Lambda$CDM model, the accelerated expansion of the universe is caused by a dark energy component that behaves like a cosmological constant, but alternative models have been proposed and are still allowed by data.

Dynamical models can be distinguished from the cosmological constant by considering the evolution of the equation of state of dark energy, $w = p/\varrho$, where $p$ and $\varrho$ are the pressure density and energy density of the fluid, respectively. 
In the cosmological constant model, $w=-1$, while for dynamical models $w=w(a)$. It is the useful to consider a Taylor expansion of the equation of state~\citep{Linder:2003}:
\begin{align}
w(a) = w_0 + w_a (1-a) \, ;
\end{align}
in the $\Lambda$CDM model we have  $w_0=-1$ and $w_a=0$. If a deviation from these values will be detected (in particular if $w_a \neq 0$), then this would suggest that the correct model is one where the dark energy component of the universe is evolving with time.

\begin{figure*}[htb!]
\center
\includegraphics[width=0.77\linewidth]{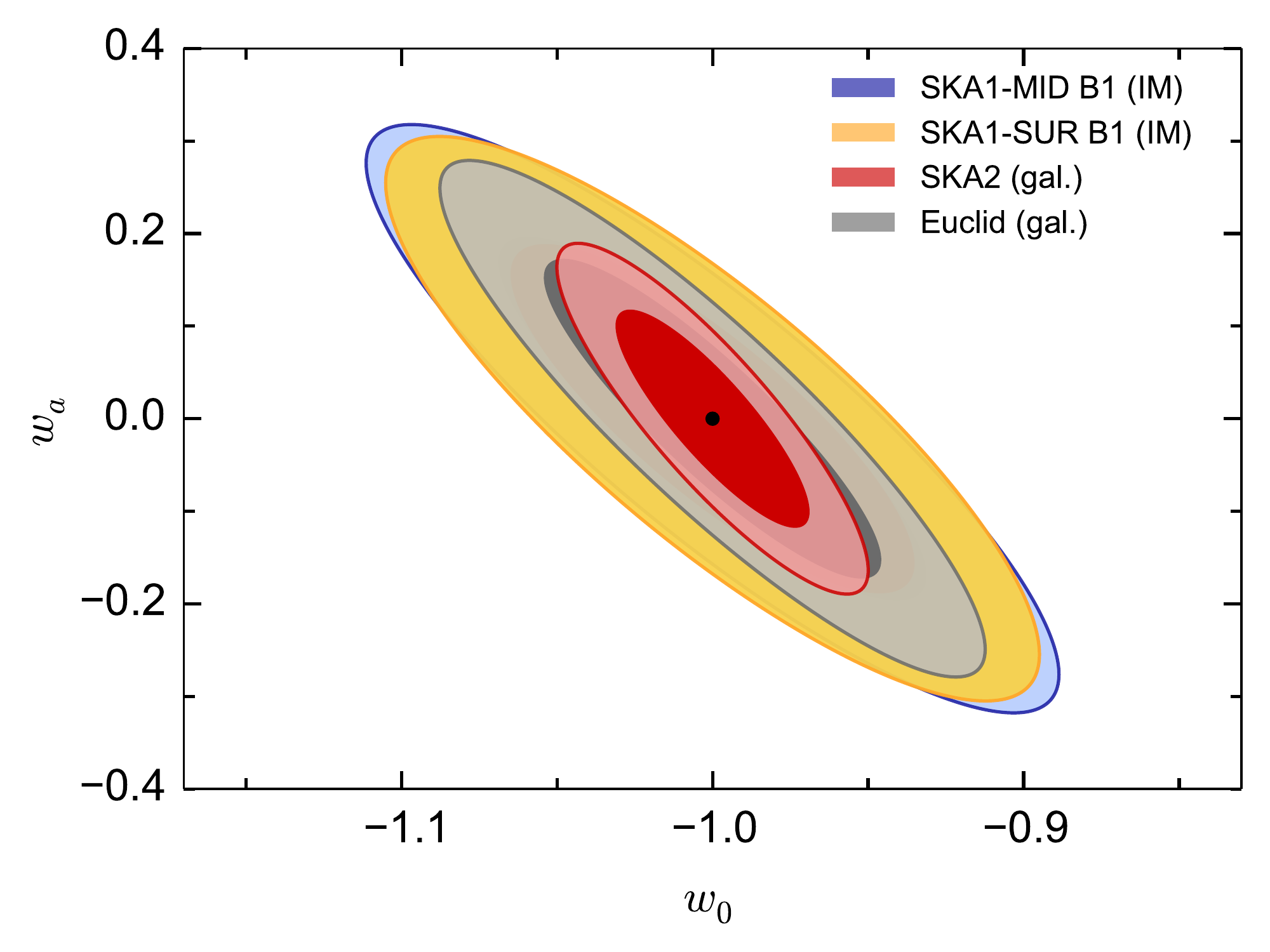}
\caption{
Predicted constraints from SKA on dynamical dark energy parameters. We show predicted constraints from SKA IM and SKA2, compared with predictions for Euclid.
}
\label{fig:w0wa}
\end{figure*}

In Figure~\ref{fig:w0wa} we plot constraints on the parameters $\{w_0, w_a\}$ (including Planck+BOSS priors), for the different SKA1 and SKA2 surveys, comparing results with predictions for the Euclid experiment.

SKA1 HI surveys will  not be able to provide competitive constraints on these parameters, so we don't show them, but results from the IM surveys will be competitive, and the SKA2 galaxy survey should be able to allow improvements on measurements of dynamical dark energy parameters over the predicted Euclid galaxy survey.

\subsubsection{Modified Growth Models}
\label{sec:mg}
Measuring the matter velocity field at the locations of the galaxies gives an 
unbiased measurement of $f\sigma_{8 \rm m}$, provided that the distribution of galaxies randomly samples matter velocities, where $f$ is given by Equation~\ref{eq:fdef} and $\sigma_{8 \rm m}$ quantifies the amplitude of fluctuations in the
matter density field.
The growth factor is sometimes parameterized as~\citep{Linder:2005}:
\begin{equation}
\label{eq:growthlinder}
D(a) = a \mbox{ {\rm exp}} \left[ \int_0^a \left[ \Omega_m^{\gamma}(a')-1 \right] \frac{d a'}{a'} \right],
\end{equation}
which leads to the following expression for $f$:
\begin{equation}
\label{eq:fgamma}
f = \left[\Omega_m(a) \right]^\gamma \, , 
\end{equation}
where:
\begin{equation}
\Omega_m(a) = \frac{\Omega_m a^{-3}}{\sum_i \Omega_i \mbox{ {\rm exp}} \left[ 3 \int_a^1 \left[ w_i(a')+1 \right] \frac{d a'}{a'} \right]},
\end{equation}
where the summation index goes over all the components of the Universe (i.e. dark matter, dark energy, curvature, radiation).
Within this formalism, $\gamma$ is a parameter that is different for different cosmological models: for example, in the standard $\Lambda$CDM+GR model it is constant, $\gamma \approx 0.55$, while it is $\approx 0.68$ for the self-accelerating DGP model~\citep[see e.g.][]{Linder:2005}. In some other cases, it is a function of the cosmological parameters or redshift~\citep[see e.g.][]{Raccanelligrowth}. 
It should be noted, however, that the parameterization given by Equation~\ref{eq:growthlinder} does not necessarily describe the growth rate in non-standard cosmologies~\citep[see e.g.][]{Schmidt:2009}.  

%\textbf{*** a bit more on different models, $f\sigma_{8 \rm m}$, $\gamma$ ***}.

In this paragraph we study how the SKA will be able to constrain the growth of structures in two different cases; we investigate constraints on the growth of structures for models with:
\begin{itemize}
\item \textbf{Parameterized growth}: \\
\vspace{0.1cm}
$\{ \rm h, \Omega_{\lambda}, \Omega_{K}, \Omega_m, \Omega_b, n_s, w_0, w_a, \sigma_8, b, \gamma, f_{\rm NL} \}$ \\
For the parameterized case we assume a model for the redshift evolution of $f$, the bias and $\sigma_8$ (and assume we have some independent measurements of them from e.g. the CMB), and we constrain the growth rate parameter $\gamma$.
Our results show that, while predictions for SKA1 HI galaxies are not competitive, the IM case is competitive with current optical surveys and comparable to future surveys such as Euclid. Constraints coming from the SKA2 galaxy survey are predicted to considerably improve forecasted results from Euclid.

In Figure~\ref{fig:w0_gamma} we plot constraints on the growth rate parameter $\gamma$ and the effective $w_0$ (including Planck+BOSS priors) for some SKA1 IM surveys and for the SKA2, comparing results with predictions for Euclid.

\begin{figure*}[htb!]
\center
\includegraphics[width=0.77\linewidth]{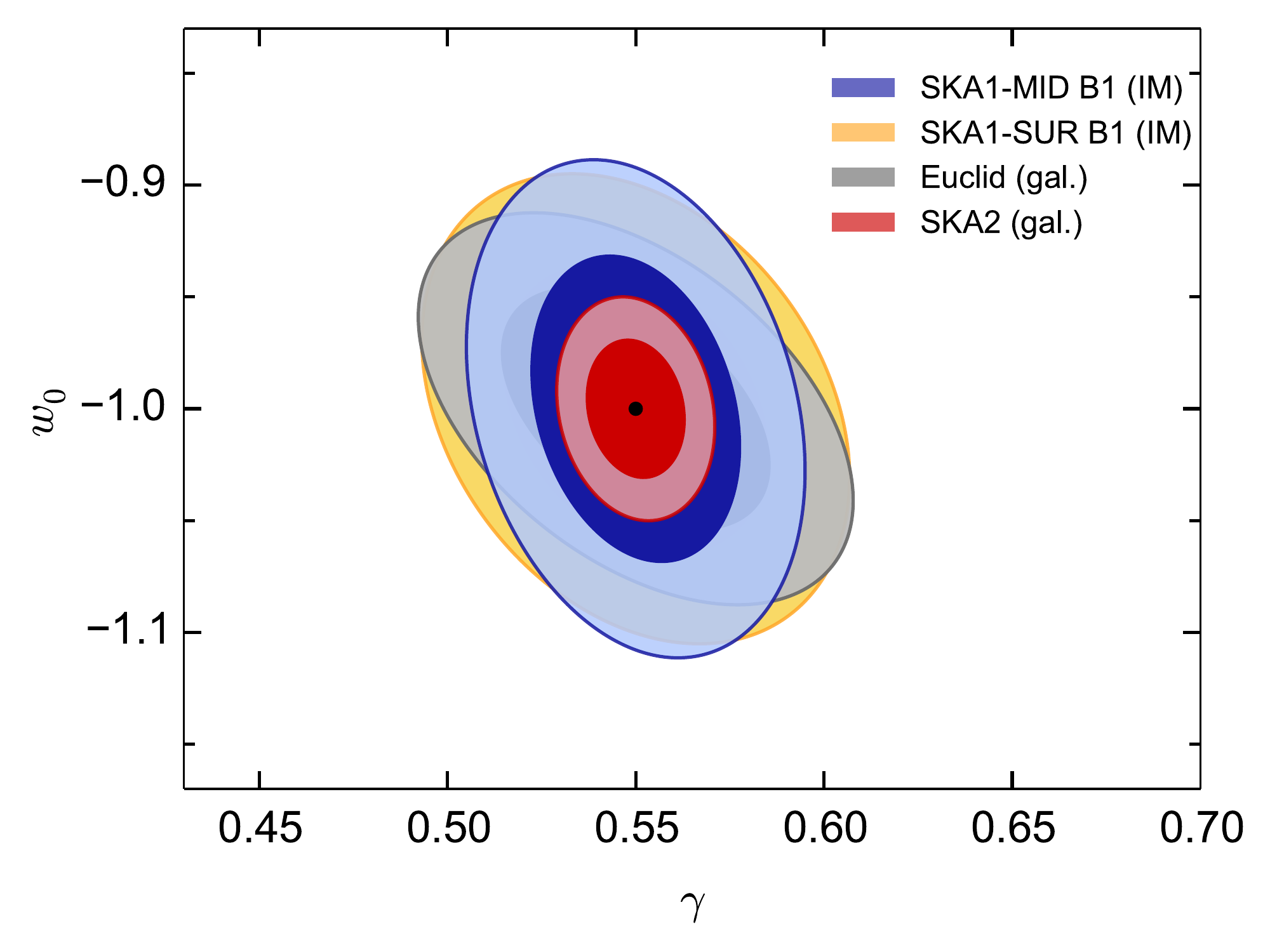}
\caption{
Predicted constraints from SKA on parameterized growth. We show predicted constraints from SKA IM and the SKA2, compared with predictions for Euclid.
}
\label{fig:w0_gamma}
\end{figure*}

\vspace{0.1cm}
\item \textbf{Unparameterized growth function}: \\
\vspace{0.1cm}
$\{ \rm h, \Omega_{\lambda}, \Omega_{K}, \Omega_m, \Omega_b, n_s, w_0, w_a, f \sigma_8, b\sigma_8 \}$ \\
In this case we assume we have no prior knowledge on the above parameters, and we constrain the combination $\{f \sigma_8, b\sigma_8 \}$.

Again our predictions show that SKA1 HI galaxy surveys are not competitive, while the IM case is competitive with current optical surveys and comparable to future surveys such as Euclid, at low redshift. Constraints coming from the SKA2 galaxy survey are predicted to be the best ones at low-z and comparable to Euclid at medium z.
This can be seen in Figure~\ref{fig:fsigma8}, where we plot constraints on the fractional precision on measurements of $f \sigma_8$ in different redshift bins.

\begin{figure*}[htb!]
\center
\includegraphics[width=0.77\linewidth]{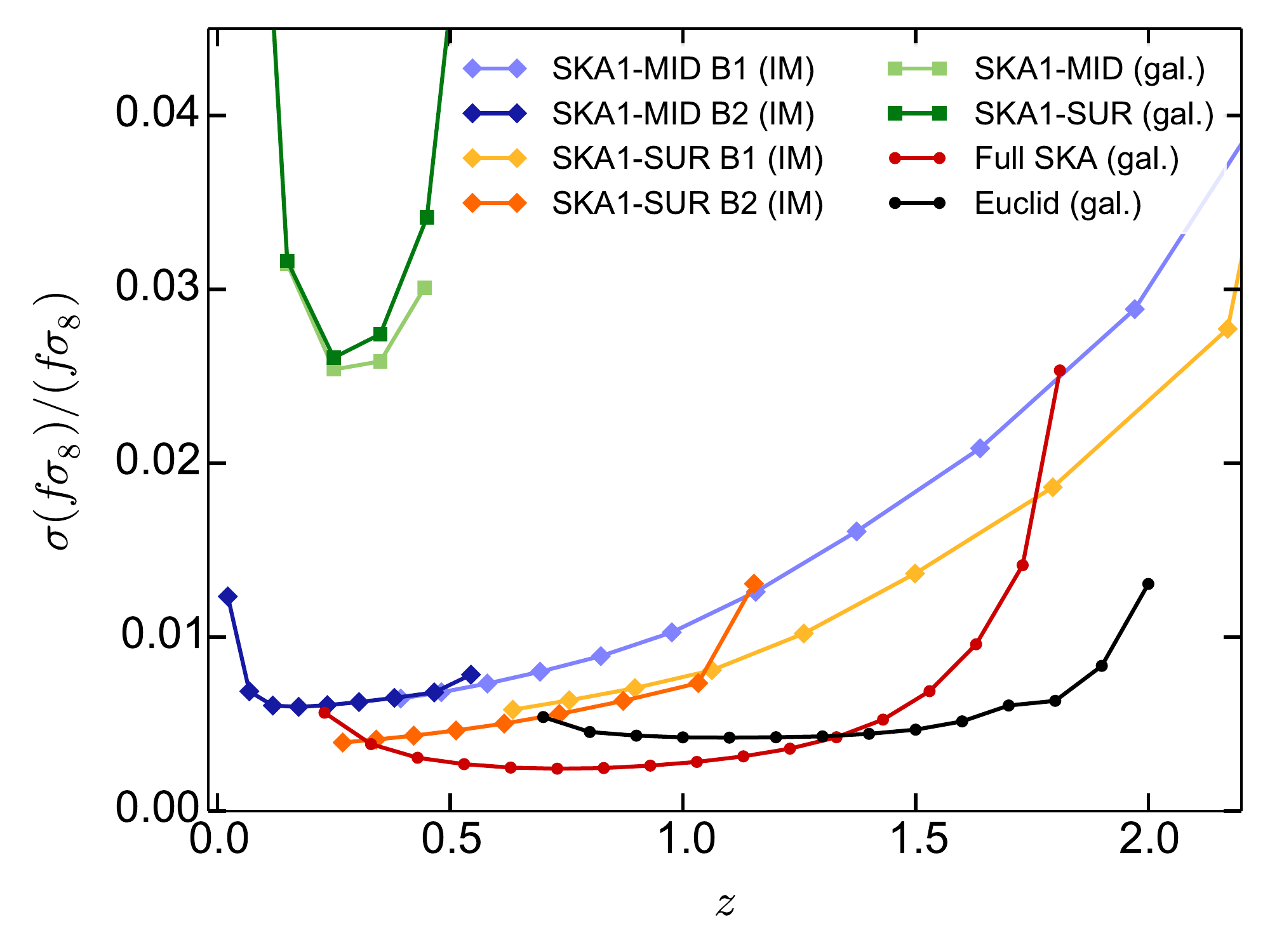}
\caption{
Predicted constraints from SKA on $f \sigma_8$ from the SKA1 (galaxy and IM) and the SKA2, compared with predicted constraints coming from the Euclid galaxy survey.
}
\label{fig:fsigma8}
\end{figure*}
\end{itemize}

\section{Discussion}
\label{sec:discussion}
In this Chapter we presented forecasts for the measurements on the growth that will be possible to obtain by measuring the full shape of the galaxy power spectrum with the SKA. 
We investigated how the different proposed SKA1 (both in the IM and HI cases) and SKA2 surveys will enable measurements of parameters describing models for the growth of structures.

In all cases analyzed, our results show that the SKA1 HI surveys will not be competitive with future galaxy surveys on the same time-scale. However, the IM case will give constraints, at low-z, at the same level of constraints provided from Euclid at medium-z.
SKA2, on the other hand, should provide the best constraints on low-redshift, and constraints that are comparable to the predicted Euclid ones up to redshift $\sim$1.5.

In this work we haven't considered systematic effects that could bias the measurements and decrease their precision; on the other hand, in our results we haven't included a proper modeling of non-linear effects, which would allow including many more modes, nor a modeling of the ultra-large-scale effects. Both would improve the constraining power of galaxy clustering measurements. Further improvements could be enabled by the use of the so-called multi-tracer technique.
Another technique that will be worth investigating derives from measuring peculiar velocities of galaxies using the Tully-Fisher relation.

Overall our results show that the SKA promises to provide the best constraints on models for the growth for the next generation of galaxy surveys.

\vspace{0.5cm}

\noindent{\bf Acknowledgments:}\\
AR is supported by the Templeton Foundation.
Part of the research described in this paper was carried out at the Jet Propulsion Laboratory, California Institute of Technology, under a contract with the National Aeronautics and Space Administration.
MV is supported by ERC StG "cosmoIGM".
MS and RM are supported by the South African SKA Project and the National Research Foundation. DB, RM and GBZ are  supported by the UK Science \& Technology Facilities Council (grant ST/K0090X/1)
GBZ is supported by Strategic Priority Research Program ``The Emergence of Cosmological Structures'' of the Chinese Academy of Sciences, Grant No. XDB09000000, by the 1000 Young Talents program in China, and by the 973 Program grant No. 2013CB837900, NSFC grant No. 11261140641, and CAS grant No. KJZD-EW-T01.

\bibliographystyle{apj}
\bibliography{SKA_RSD}

\begin{thebibliography}{90}
\expandafter\ifx\csname natexlab\endcsname\relax\def\natexlab#1{#1}\fi

\bibitem[{Anselmi {et~al.}(2010)Anselmi, Matarrese, \& Pietroni}]{Anselmi:2010}
Anselmi, S., Matarrese, S., \& Pietroni, M. 2010, arXiv:1011.4477

\bibitem[{Anselmi \& Pietroni(2012)}]{Anselmi:2012}
Anselmi, S. \& Pietroni, M. 2012, arXiv:1205.2235

\bibitem[{Baker {et~al.}(2011)Baker, Ferreira, Skordis, \& Zuntz}]{Baker:2011}
Baker, T., Ferreira, P.~G., Skordis, C., \& Zuntz, J. 2011, Phys. Rev. D, 84,
  124018

\bibitem[{Baldauf {et~al.}(2011)Baldauf, Seljak, Senatore, \&
  Zaldarriaga}]{Baldauf:2011}
Baldauf, T., Seljak, U., Senatore, L., \& Zaldarriaga, M. 2011, arXiv:1106.5507

\bibitem[{Ballinger {et~al.}(1996)Ballinger, Peacock, \&
  Heavens}]{Ballinger:1996}
Ballinger, W.~E., Peacock, J.~A., \& Heavens, A.~F. 1996, astro-ph/9605017

\bibitem[{Bernardeau {et~al.}(2001)Bernardeau, Colombi, Gazta{\~n}aga, \&
  Scoccimarro}]{Bernardeau:2002}
Bernardeau, F., Colombi, S., Gazta{\~n}aga, E., \& Scoccimarro, R. 2001,
  arXiv.org, 1

\bibitem[{Bertacca {et~al.}(2012)Bertacca, Maartens, Raccanelli, \&
  Clarkson}]{Bertacca:2012}
Bertacca, D., Maartens, R., Raccanelli, A., \& Clarkson, C. 2012,
  JCAP10(2012)025

\bibitem[{Blake {et~al.}(2013)Blake, Baldry, Bland-Hawthorn, Christodoulou,
  Colless, Conselice, Driver, Hopkins, Liske, Loveday, Norberg, Peacock, Poole,
  \& Robotham}]{Blake:2013}
Blake, C., Baldry, I.~K., Bland-Hawthorn, J., Christodoulou, L., Colless, M.,
  Conselice, C.~J., Driver, S.~P., Hopkins, A.~M., Liske, J., Loveday, J.,
  Norberg, P., Peacock, J.~A., Poole, G.~B., \& Robotham, A.~S. 2013,
  arXiv:1309.5556

\bibitem[{Blake {et~al.}(2012)Blake, Brough, Colless, Contreras, Couch, Croom,
  Croton, Davis, Drinkwater, Forster, Gilbank, Gladders, Glazebrook, Jelliffe,
  Jurek, hui Li, Madore, Martin, Pimbblet, Poole, Pracy, Sharp, Wisnioski,
  Woods, Wyder, \& Yee}]{Blake:2012}
Blake, C., Brough, S., Colless, M., Contreras, C., Couch, W., Croom, S.,
  Croton, D., Davis, T., Drinkwater, M.~J., Forster, K., Gilbank, D., Gladders,
  M., Glazebrook, K., Jelliffe, B., Jurek, R.~J., hui Li, I., Madore, B.,
  Martin, C., Pimbblet, K., Poole, G.~B., Pracy, M., Sharp, R., Wisnioski, E.,
  Woods, D., Wyder, T., \& Yee, H. 2012, arXiv:1204.3674

\bibitem[{Blake {et~al.}(2011)Blake, Brough, Colless, Contreras, Couch, Croom,
  Davis, Drinkwater, Forster, Gilbank, Gladders, Glazebrook, Jelliffe, Jurek,
  hui Li, Madore, Martin, Pimbblet, Poole, Pracy, Sharp, Wisnioski, Woods,
  Wyder, \& Yee}]{Blake:2011}
Blake, C., Brough, S., Colless, M., Contreras, C., Couch, W., Croom, S., Davis,
  T., Drinkwater, M.~J., Forster, K., Gilbank, D., Gladders, M., Glazebrook,
  K., Jelliffe, B., Jurek, R.~J., hui Li, I., Madore, B., Martin, C., Pimbblet,
  K., Poole, G., Pracy, M., Sharp, R., Wisnioski, E., Woods, D., Wyder, T., \&
  Yee, H. 2011, arXiv:1104.2948

\bibitem[{Blake {et~al.}(2010)Blake, Brough, Colless, Couch, Croom, Davis,
  Drinkwater, Forster, Glazebrook, Jelliffe, Jurek, hui Li, Madore, Martin,
  Pimbblet, Poole, Pracy, Sharp, Wisnioski, Woods, \& Wyder}]{Blake:2010}
Blake, C., Brough, S., Colless, M., Couch, W., Croom, S., Davis, T.,
  Drinkwater, M.~J., Forster, K., Glazebrook, K., Jelliffe, B., Jurek, R.~J.,
  hui Li, I., Madore, B., Martin, C., Pimbblet, K., Poole, G.~B., Pracy, M.,
  Sharp, R., Wisnioski, E., Woods, D., \& Wyder, T. 2010, arXiv:1003.5721

\bibitem[{Bonvin \& Durrer(2011)}]{Bonvin:2011}
Bonvin, C. \& Durrer, R. 2011, Phys.Rev.D, 84, 063505

\bibitem[{Brans(2000)}]{Brans:2000}
Brans, C.~H. 2000, arXiv:0506063

\bibitem[{Bull {et~al.}(2015)Bull, Camera, Raccanelli, Blake, Ferreira, Santos,
  \& Schwarz}]{Bull}
Bull, P., Camera, S., Raccanelli, A., Blake, C., Ferreira, P.~G., Santos,
  M.~G., \& Schwarz, D.~J. 2015, Proceedings of ``Advancing Astrophysics with
  the Square Kilometre Array'', PoS (AASKA15)

\bibitem[{Bull {et~al.}(2014)Bull, Ferreira, Patel, \& Santos}]{Bull:2014rha}
Bull, P., Ferreira, P.~G., Patel, P., \& Santos, M.~G. 2014, arXiv:1405.1452

\bibitem[{Camera {et~al.}(2015)Camera, Raccanelli, Bull, Bertacca, Chen,
  Ferreira, Kunz, Maartens, Mao, Santos, Shapiro, Viel, \& Xu}]{Camera}
Camera, S., Raccanelli, A., Bull, P., Bertacca, D., Chen, X., Ferreira, P.~G.,
  Kunz, M., Maartens, R., Mao, Y., Santos, M.~G., Shapiro, P.~R., Viel, M., \&
  Xu, Y. 2015, Proceedings of ``Advancing Astrophysics with the Square
  Kilometre Array'', PoS (AASKA15)

\bibitem[{Camera {et~al.}(2012)Camera, Santos, Bacon, Jarvis, McAlpine, Norris,
  Raccanelli, \& R{\"o}ttgering}]{Camera:2012ly}
Camera, S., Santos, M.~G., Bacon, D.~J., Jarvis, M.~J., McAlpine, K., Norris,
  R.~P., Raccanelli, A., \& R{\"o}ttgering, H. 2012, MNRAS, 427, 2079

\bibitem[{Carilli \& Rawlings(2004)}]{Carilli:2004}
Carilli, C.~L. \& Rawlings, S. 2004, New Astron.Rev.

\bibitem[{Carroll {et~al.}(2004)Carroll, Duvvuri, Trodden, \&
  Turner}]{Carroll:2004}
Carroll, S.~M., Duvvuri, V., Trodden, M., \& Turner, M.~S. 2004, Phys.Rev.D,
  70, 043528

\bibitem[{Carron \& Szapudi(2013)}]{Carron:2013}
Carron, J. \& Szapudi, I. 2013, arXiv: 1306.1230

\bibitem[{Challinor \& Lewis(2011)}]{Challinor:2011}
Challinor, A. \& Lewis, A. 2011, Phys.Rev.D, 84, 043516

\bibitem[{Crocce {et~al.}(2013)Crocce, Castander, Gaztanaga, Fosalba, \&
  Carretero}]{Crocce:2013}
Crocce, M., Castander, F.~J., Gaztanaga, E., Fosalba, P., \& Carretero, J.
  2013, arXiv:1312.2013

\bibitem[{Cyr-Racine {et~al.}(2014)Cyr-Racine, de~Putter, Raccanelli, \&
  Sigurdson}]{Cyr-Racine:2014}
Cyr-Racine, F.-Y., de~Putter, R., Raccanelli, A., \& Sigurdson, K. 2014, Phys.
  Rev. D, 89, 063517

\bibitem[{de~Putter {et~al.}(2012)de~Putter, Mena, Giusarma, Ho, Cuesta, Seo,
  Ross, White, Bizyaev, Brewington, Kirkby, Malanushenko, Malanushenko,
  Oravetz, Pan, Percival, Ross, Schneider, Shelden, Simmons, \&
  Snedden}]{dePutter:2012}
de~Putter, R., Mena, O., Giusarma, E., Ho, S., Cuesta, A., Seo, H.-J., Ross,
  A., White, M., Bizyaev, D., Brewington, H., Kirkby, D., Malanushenko, E.,
  Malanushenko, V., Oravetz, D., Pan, K., Percival, W.~J., Ross, N.~P.,
  Schneider, D.~P., Shelden, A., Simmons, A., \& Snedden, S. 2012,
  arXiv:1201.1909

\bibitem[{Dewdney {et~al.}(2009)Dewdney, Hall, Schilizzi, \&
  Lazio}]{Dewdney:2009}
Dewdney, P., Hall, P., Schilizzi, R., \& Lazio, J. 2009, Proceedings of the
  IEEE, 97

\bibitem[{Dio {et~al.}(2014)Dio, Montanari, Durrer, \&
  Lesgourgues}]{DiDio:2014}
Dio, E.~D., Montanari, F., Durrer, R., \& Lesgourgues, J. 2014, JCAP, 01, 042

\bibitem[{Dvali {et~al.}(2000)Dvali, Gabadadze, \& Porrati}]{Dvali:2000}
Dvali, G., Gabadadze, G., \& Porrati, M. 2000, Phys.Lett., B484, 112

\bibitem[{Dvorkin {et~al.}(2014)Dvorkin, Blum, \& Kamionkowski}]{Dvorkin:2014}
Dvorkin, C., Blum, K., \& Kamionkowski, M. 2014, Phys.Rev.D, 89, 023519

\bibitem[{Fisher(1935)}]{Fisher:1935}
Fisher, R.~A. 1935, J. Roy. Stat. Soc., 98

\bibitem[{{Freudling} {et~al.}(2011){Freudling}, {Staveley-Smith}, {Catinella},
  {Minchin}, {Calabretta}, {Momjian}, {Zwaan}, {Meyer}, \& {O'Neil}}]{Arecibo}
{Freudling}, W., {Staveley-Smith}, L., {Catinella}, B., {Minchin}, R.,
  {Calabretta}, M., {Momjian}, E., {Zwaan}, M., {Meyer}, M., \& {O'Neil}, K.
  2011, APJ, 727, 40

\bibitem[{Guzzo {et~al.}(2008)Guzzo, Pierleoni, Meneux, Branchini, F{\`e}vre,
  Marinoni, Garilli, Blaizot, Lucia, Pollo, McCracken, Bottini, Brun, Maccagni,
  Picat, Scaramella, Scodeggio, Tresse, Vettolani, Zanichelli, Adami, Arnouts,
  Bardelli, Bolzonella, Bongiorno, Cappi, Charlot, Ciliegi, Contini, Cucciati,
  de~la Torre, Dolag, Foucaud, Franzetti, Gavignaud, Ilbert, Iovino,
  Lamareille, Marano, Mazure, Memeo, Merighi, Moscardini, Paltani, Pell{\`o},
  Perez-Montero, Pozzetti, Radovich, Vergani, Zamorani, \& Zucca}]{Guzzo:2008}
Guzzo, L., Pierleoni, M., Meneux, B., Branchini, E., F{\`e}vre, O.~L.,
  Marinoni, C., Garilli, B., Blaizot, J., Lucia, G.~D., Pollo, A., McCracken,
  H.~J., Bottini, D., Brun, V.~L., Maccagni, D., Picat, J.~P., Scaramella, R.,
  Scodeggio, M., Tresse, L., Vettolani, G., Zanichelli, A., Adami, C., Arnouts,
  S., Bardelli, S., Bolzonella, M., Bongiorno, A., Cappi, A., Charlot, S.,
  Ciliegi, P., Contini, T., Cucciati, O., de~la Torre, S., Dolag, K., Foucaud,
  S., Franzetti, P., Gavignaud, I., Ilbert, O., Iovino, A., Lamareille, F.,
  Marano, B., Mazure, A., Memeo, P., Merighi, R., Moscardini, L., Paltani, S.,
  Pell{\`o}, R., Perez-Montero, E., Pozzetti, L., Radovich, M., Vergani, D.,
  Zamorani, G., \& Zucca, E. 2008, Nature, 451, 541

\bibitem[{Hamilton(1997)}]{Hamilton:1997}
Hamilton, A. J.~S. 1997, astro-ph/9708102, astro-ph

\bibitem[{Jackson(1972)}]{Jackson:1972}
Jackson, C.~J. 1972, MNRAS, 156

\bibitem[{Jain \& Zhang(2008)}]{Jain:2007}
Jain, B. \& Zhang, P. 2008, Phys.Rev.D, 78, 063503

\bibitem[{Jennings(2012)}]{Jennings:2012}
Jennings, E. 2012, 2012MNRAS.427L..25J

\bibitem[{Jeong {et~al.}(2011)Jeong, Schmidt, \& Hirata}]{Jeong:2011}
Jeong, D., Schmidt, F., \& Hirata, C.~M. 2011, 2012, PRD 85, 023504

\bibitem[{Johnston {et~al.}(2008)Johnston, Taylor, Bailes, Bartel, Baugh,
  Bietenholz, Blake, Braun, Brown, Chatterjee, Darling, Deller, Dodson,
  Edwards, Ekers, Ellingsen, Feain, Gaensler, Haverkorn, Hobbs, Hopkins,
  Jackson, James, Joncas, Kaspi, Kilborn, Koribalski, Kothes, Landecker, Lenc,
  Lovell, Macquart, Manchester, Matthews, McClure-Griffiths, Norris, Pen,
  Phillips, Power, Protheroe, Sadler, Schmidt, Stairs, Staveley-Smith, Stil,
  Tingay, Tzioumis, Walker, Wall, \& Wolleben}]{askap}
Johnston, S., Taylor, R., Bailes, M., Bartel, N., Baugh, C., Bietenholz, M.,
  Blake, C., Braun, R., Brown, J., Chatterjee, S., Darling, J., Deller, A.,
  Dodson, R., Edwards, P., Ekers, R., Ellingsen, S., Feain, I., Gaensler, B.,
  Haverkorn, M., Hobbs, G., Hopkins, A., Jackson, C., James, C., Joncas, G.,
  Kaspi, V., Kilborn, V., Koribalski, B., Kothes, R., Landecker, T., Lenc, A.,
  Lovell, J., Macquart, J.-P., Manchester, R., Matthews, D., McClure-Griffiths,
  N., Norris, R., Pen, U.-L., Phillips, C., Power, C., Protheroe, R., Sadler,
  E., Schmidt, B., Stairs, I., Staveley-Smith, L., Stil, J., Tingay, S.,
  Tzioumis, A., Walker, M., Wall, J., \& Wolleben, M. 2008, arXiv:0810.5187

\bibitem[{Kaiser(1984)}]{Kaiser:1984}
Kaiser, N. 1984, Astrophysical Journal, 284, L9

\bibitem[{Kaiser(1987)}]{Kaiser:1987}
---. 1987, Royal Astronomical Society, 227, 1

\bibitem[{Kwan {et~al.}(2011)Kwan, Lewis, \& Linder}]{Kwan:2011}
Kwan, J., Lewis, G.~F., \& Linder, E.~V. 2011, ApJ, 2012, 748, 78

\bibitem[{Linder(2003)}]{Linder:2003}
Linder, E.~V. 2003, Physical Review Letters, 90, 91301

\bibitem[{Linder(2005)}]{Linder:2005}
---. 2005, astro-ph/0507263, astro-ph

\bibitem[{Linder(2007)}]{Linder:2007}
---. 2007, arXiv:0709.1113, astro-ph

\bibitem[{Matsubara(1999)}]{Matsubara:1999}
Matsubara, T. 1999, astro-ph/9908056

\bibitem[{Montanari \& Durrer(2012)}]{Montanari:2012}
Montanari, F. \& Durrer, R. 2012, Phys. Rev., D86, 063503

\bibitem[{Musso {et~al.}(2012)Musso, Paranjape, \& Sheth}]{Musso:2012}
Musso, M., Paranjape, A., \& Sheth, R.~K. 2012, Monthly Notices of the Royal
  Astronomical Society, 427, 3145

\bibitem[{Neyrinck {et~al.}(2009)Neyrinck, Szapudi, \& Szalay}]{Neyrinck:2009}
Neyrinck, M.~C., Szapudi, I., \& Szalay, A.~S. 2009, Astrophys.J., 698, L90

\bibitem[{Neyrinck {et~al.}(2011)Neyrinck, Szapudi, \& Szalay}]{Neyrinck:2011}
---. 2011, Astrophys. J., 731, 116

\bibitem[{Nolta {et~al.}(2004)Nolta, Wright, Page, Bennett, Halpern, Hinshaw,
  Jarosik, Kogut, Limon, Meyer, Spergel, Tucker, \& Wollack}]{Nolta:2004}
Nolta, M.~R., Wright, E.~L., Page, L., Bennett, C.~L., Halpern, M., Hinshaw,
  G., Jarosik, N., Kogut, A., Limon, M., Meyer, S.~S., Spergel, D.~N., Tucker,
  G.~S., \& Wollack, E. 2004, Astrophys.J., 608, 10

\bibitem[{Papai \& Szapudi(2008)}]{Papai:2008}
Papai, P. \& Szapudi, I. 2008, arXiv:0802.2940, astro-ph

\bibitem[{Peacock {et~al.}(2001)Peacock, Colless, Peacock, Baugh,
  Bland-Hawthorn, Bridges, Cannon, Cole, Collins, Couch, Cross, Dalton, Deeley,
  Propris, Driver, Efstathiou, Ellis, Frenk, Glazebrook, Jackson, Lahav, Lewis,
  Lumsden, Maddox, Madgwick, Norberg, Percival, Peterson, Sutherland, \&
  Taylor}]{Peacock:2001}
Peacock, J.~A., Colless, M., Peacock, J., Baugh, C.~M., Bland-Hawthorn, J.,
  Bridges, T., Cannon, R., Cole, S., Collins, C., Couch, W., Cross, N., Dalton,
  G., Deeley, K., Propris, R.~D., Driver, S., Efstathiou, G., Ellis, R.~S.,
  Frenk, C.~S., Glazebrook, K., Jackson, C., Lahav, O., Lewis, I., Lumsden, S.,
  Maddox, S., Madgwick, D., Norberg, P., Percival, W., Peterson, B.,
  Sutherland, W., \& Taylor, K. 2001, Deep Fields: Proceedings of the ESO
  Workshop Held at Garching, 221, iSBN: 3-540-42799-6 (c) 2001: Springer-Verlag

\bibitem[{Percival {et~al.}(2004)Percival, Burkey, Heavens, Taylor, Cole,
  Peacock, Baugh, Bland-Hawthorn, Bridges, Cannon, Colless, Collins, Couch,
  Dalton, Propris, Driver, Efstathiou, Ellis, Frenk, Glazebrook, Jackson,
  Lahav, Lewis, Lumsden, Maddox, Norberg, Peterson, Sutherland, \&
  Taylor}]{Percival:2004}
Percival, W.~J., Burkey, D., Heavens, A., Taylor, A., Cole, S., Peacock, J.~A.,
  Baugh, C.~M., Bland-Hawthorn, J., Bridges, T., Cannon, R., Colless, M.,
  Collins, C., Couch, W., Dalton, G., Propris, R.~D., Driver, S.~P.,
  Efstathiou, G., Ellis, R.~S., Frenk, C.~S., Glazebrook, K., Jackson, C.,
  Lahav, O., Lewis, I., Lumsden, S., Maddox, S., Norberg, P., Peterson, B.~A.,
  Sutherland, W., \& Taylor, K. 2004, Monthly Notices of the Royal Astronomical
  Society, 353, 1201

\bibitem[{Raccanelli {et~al.}(2013{\natexlab{a}})Raccanelli, Bertacca, Dore, \&
  Maartens}]{Raccanelli:2013}
Raccanelli, A., Bertacca, D., Dore, O., \& Maartens, R. 2013{\natexlab{a}},
  arXiv:1306.6646

\bibitem[{Raccanelli {et~al.}(2013{\natexlab{b}})Raccanelli, Bertacca,
  Maartens, Clarkson, \& Dor{\'e}}]{Raccanelliradial}
Raccanelli, A., Bertacca, D., Maartens, R., Clarkson, C., \& Dor{\'e}, O.
  2013{\natexlab{b}}, arXiv:1311.6813

\bibitem[{{Raccanelli} {et~al.}(2013){Raccanelli}, {Bertacca}, {Pietrobon},
  {Schmidt}, {Samushia}, {Bartolo}, {Dor{\'e}}, {Matarrese}, \&
  {Percival}}]{Raccanelligrowth}
{Raccanelli}, A., {Bertacca}, D., {Pietrobon}, D., {Schmidt}, F., {Samushia},
  L., {Bartolo}, N., {Dor{\'e}}, O., {Matarrese}, S., \& {Percival}, W.~J.
  2013, \mnras

\bibitem[{Raccanelli {et~al.}(2008)Raccanelli, Bonaldi, Negrello, Matarrese,
  Tormen, \& Zotti}]{Raccanelli:2008nx}
Raccanelli, A., Bonaldi, A., Negrello, M., Matarrese, S., Tormen, G., \& Zotti,
  G.~D. 2008, Mon.Not.Roy.Astron.Soc., 386, 2161

\bibitem[{Raccanelli {et~al.}(2014)Raccanelli, Dor{\'e}, Bacon, Maartens,
  Santos, Camera, Davis, Drinkwater, Jarvis, Norris, \&
  Parkinson}]{Raccanelli:2014isw}
Raccanelli, A., Dor{\'e}, O., Bacon, D.~J., Maartens, R., Santos, M.~G.,
  Camera, S., Davis, T., Drinkwater, M.~J., Jarvis, M., Norris, R., \&
  Parkinson, D. 2014, arXiv:1406.0010

\bibitem[{{Raccanelli} {et~al.}(2010){Raccanelli}, {Samushia}, \&
  {Percival}}]{Raccanelli:2010fk}
{Raccanelli}, A., {Samushia}, L., \& {Percival}, W.~J. 2010, \mnras, 409, 1525

\bibitem[{Raccanelli {et~al.}(2012)Raccanelli, Zhao, Bacon, Jarvis, Percival,
  Norris, Rottgering, Abdalla, Cress, Kubwimana, Lindsay, Nichol, Santos, \&
  Schwarz}]{Raccanelliradio}
Raccanelli, A., Zhao, G.-B., Bacon, D.~J., Jarvis, M.~J., Percival, W.~J.,
  Norris, R.~P., Rottgering, H., Abdalla, F.~B., Cress, C.~M., Kubwimana,
  J.-C., Lindsay, S., Nichol, R.~C., Santos, M.~G., \& Schwarz, D.~J. 2012,
  MNRAS, 424

\bibitem[{{Reid} {et~al.}(2010){Reid}, {Percival}, {Eisenstein}, {Verde},
  {Spergel}, {Skibba}, {Bahcall}, {Budavari}, {Frieman}, {Fukugita}, {Gott},
  {Gunn}, {Ivezi{\'c}}, {Knapp}, {Kron}, {Lupton}, {McKay}, {Meiksin},
  {Nichol}, {Pope}, {Schlegel}, {Schneider}, {Stoughton}, {Strauss}, {Szalay},
  {Tegmark}, {Vogeley}, {Weinberg}, {York}, \& {Zehavi}}]{Reid:2010}
{Reid}, B.~A., {Percival}, W.~J., {Eisenstein}, D.~J., {Verde}, L., {Spergel},
  D.~N., {Skibba}, R.~A., {Bahcall}, N.~A., {Budavari}, T., {Frieman}, J.~A.,
  {Fukugita}, M., {Gott}, J.~R., {Gunn}, J.~E., {Ivezi{\'c}}, {\v Z}., {Knapp},
  G.~R., {Kron}, R.~G., {Lupton}, R.~H., {McKay}, T.~A., {Meiksin}, A.,
  {Nichol}, R.~C., {Pope}, A.~C., {Schlegel}, D.~J., {Schneider}, D.~P.,
  {Stoughton}, C., {Strauss}, M.~A., {Szalay}, A.~S., {Tegmark}, M., {Vogeley},
  M.~S., {Weinberg}, D.~H., {York}, D.~G., \& {Zehavi}, I. 2010, MNRAS, 404, 60

\bibitem[{Reid {et~al.}(2012)Reid, Samushia, White, Percival, Manera,
  Padmanabhan, Ross, S{\'a}nchez, Bailey, Bizyaev, Bolton, Brewington,
  Brinkmann, Brownstein, Cuesta, Eisenstein, Gunn, Honscheid, Malanushenko,
  Malanushenko, Maraston, McBride, Muna, Nichol, Oravetz, Pan, de~Putter, Roe,
  Ross, Schlegel, Schneider, Seo, Shelden, Sheldon, Simmons, Skibba, Snedden,
  Swanson, Thomas, Tinker, Tojeiro, Verde, Wake, Weaver, Weinberg, Zehavi, \&
  Zhao}]{Reid:2012}
Reid, B.~A., Samushia, L., White, M., Percival, W.~J., Manera, M., Padmanabhan,
  N., Ross, A.~J., S{\'a}nchez, A.~G., Bailey, S., Bizyaev, D., Bolton, A.~S.,
  Brewington, H., Brinkmann, J., Brownstein, J.~R., Cuesta, A.~J., Eisenstein,
  D.~J., Gunn, J.~E., Honscheid, K., Malanushenko, E., Malanushenko, V.,
  Maraston, C., McBride, C.~K., Muna, D., Nichol, R.~C., Oravetz, D., Pan, K.,
  de~Putter, R., Roe, N.~A., Ross, N.~P., Schlegel, D.~J., Schneider, D.~P.,
  Seo, H.-J., Shelden, A., Sheldon, E.~S., Simmons, A., Skibba, R.~A., Snedden,
  S., Swanson, M. E.~C., Thomas, D., Tinker, J., Tojeiro, R., Verde, L., Wake,
  D.~A., Weaver, B.~A., Weinberg, D.~H., Zehavi, I., \& Zhao, G.-B. 2012,
  arXiv:1203.6641

\bibitem[{{Ross} {et~al.}(2013){Ross}, {Percival}, {Carnero}, {Zhao}, {Manera},
  {Raccanelli}, {Aubourg}, {Bizyaev}, {Brewington}, {Brinkmann}, {Brownstein},
  {Cuesta}, {da Costa}, {Eisenstein}, {Ebelke}, {Guo}, {Hamilton},
  {Maga{\~n}a}, {Malanushenko}, {Malanushenko}, {Maraston}, {Montesano},
  {Nichol}, {Oravetz}, {Pan}, {Prada}, {S{\'a}nchez}, {Samushia}, {Schlegel},
  {Schneider}, {Seo}, {Sheldon}, {Simmons}, {Snedden}, {Swanson}, {Thomas},
  {Tinker}, {Tojeiro}, \& {Zehavi}}]{Ross:2013}
{Ross}, A.~J., {Percival}, W.~J., {Carnero}, A., {Zhao}, G.-b., {Manera}, M.,
  {Raccanelli}, A., {Aubourg}, E., {Bizyaev}, D., {Brewington}, H.,
  {Brinkmann}, J., {Brownstein}, J.~R., {Cuesta}, A.~J., {da Costa}, L.~A.~N.,
  {Eisenstein}, D.~J., {Ebelke}, G., {Guo}, H., {Hamilton}, J.-C.,
  {Maga{\~n}a}, M.~V., {Malanushenko}, E., {Malanushenko}, V., {Maraston}, C.,
  {Montesano}, F., {Nichol}, R.~C., {Oravetz}, D., {Pan}, K., {Prada}, F.,
  {S{\'a}nchez}, A.~G., {Samushia}, L., {Schlegel}, D.~J., {Schneider}, D.~P.,
  {Seo}, H.-J., {Sheldon}, A., {Simmons}, A., {Snedden}, S., {Swanson},
  M.~E.~C., {Thomas}, D., {Tinker}, J.~L., {Tojeiro}, R., \& {Zehavi}, I. 2013,
  \mnras, 428, 1116

\bibitem[{Rottgering {et~al.}(2011)Rottgering, Afonso, Barthel, Batejat, Best,
  Bonafede, Bruggen, Brunetti, Chyzy, Conway, Gasperin, Ferrari, Haverkorn,
  Heald, Hoeft, Jackson, Jarvis, Ker, Lehnert, Macario, McKean, Miley,
  Morganti, Oosterloo, Orru, Pizzo, Rafferty, Shulevski, Tasse, van Bemmel,
  van~der Tol, van Weeren, Verheijen, White, Wise, \& Collaboration}]{lofar}
Rottgering, H., Afonso, J., Barthel, P., Batejat, F., Best, P., Bonafede, A.,
  Bruggen, M., Brunetti, G., Chyzy, K., Conway, J., Gasperin, F.~D., Ferrari,
  C., Haverkorn, M., Heald, G., Hoeft, M., Jackson, N., Jarvis, M., Ker, L.,
  Lehnert, M., Macario, G., McKean, J., Miley, G., Morganti, R., Oosterloo, T.,
  Orru, E., Pizzo, R., Rafferty, D., Shulevski, A., Tasse, C., van Bemmel, I.,
  van~der Tol, B., van Weeren, R., Verheijen, M., White, G., Wise, M., \&
  Collaboration, L. 2011, arXiv:1107.1606

\bibitem[{{Samushia} {et~al.}(2012){Samushia}, {Percival}, \&
  {Raccanelli}}]{Samushia:2011wa}
{Samushia}, L., {Percival}, W.~J., \& {Raccanelli}, A. 2012, \mnras, 420, 2102

\bibitem[{Samushia {et~al.}(2013)Samushia, Reid, White, Percival, Cuesta,
  Lombriser, Manera, Nichol, Schneider, Bizyaev, Brewington, Malanushenko,
  Malanushenko, Oravetz, Pan, Simmons, Shelden, Snedden, Tinker, Weaver, York,
  \& Zhao}]{Samushia:2013}
Samushia, L., Reid, B.~A., White, M., Percival, W.~J., Cuesta, A.~J.,
  Lombriser, L., Manera, M., Nichol, R.~C., Schneider, D.~P., Bizyaev, D.,
  Brewington, H., Malanushenko, E., Malanushenko, V., Oravetz, D., Pan, K.,
  Simmons, A., Shelden, A., Snedden, S., Tinker, J.~L., Weaver, B.~A., York,
  D.~G., \& Zhao, G.-B. 2013, Monthly Notices of the Royal Astronomical
  Society, Volume 429, Issue, 2, p.1514

\bibitem[{Samushia {et~al.}(2014)Samushia, Reid, White, Percival, Cuesta, Zhao,
  Ross, Manera, Aubourg, Beutler, Brinkmann, Brownstein, Dawson, Eisenstein,
  Ho, Honscheid, Maraston, Montesano, Nichol, Roe, Ross, S{\'a}nchez, Schlegel,
  Schneider, Streblyanska, Thomas, Tinker, Wake, Weaver, \&
  Zehavi}]{Samushia:2014}
Samushia, L., Reid, B.~A., White, M., Percival, W.~J., Cuesta, A.~J., Zhao,
  G.-B., Ross, A.~J., Manera, M., Aubourg, {\'E}., Beutler, F., Brinkmann, J.,
  Brownstein, J.~R., Dawson, K.~S., Eisenstein, D.~J., Ho, S., Honscheid, K.,
  Maraston, C., Montesano, F., Nichol, R.~C., Roe, N.~A., Ross, N.~P.,
  S{\'a}nchez, A.~G., Schlegel, D.~J., Schneider, D.~P., Streblyanska, A.,
  Thomas, D., Tinker, J.~L., Wake, D.~A., Weaver, B.~A., \& Zehavi, I. 2014,
  Monthly Notices of the Royal Astronomical Society, Volume 439, Issue, 4,
  p.3504

\bibitem[{{S{\'a}nchez} {et~al.}(2012){S{\'a}nchez}, {Sc{\'o}ccola}, {Ross},
  {Percival}, {Manera}, {Montesano}, {Mazzalay}, {Cuesta}, {Eisenstein},
  {Kazin}, {McBride}, {Mehta}, {Montero-Dorta}, {Padmanabhan}, {Prada},
  {Rubi{\~n}o-Mart{\'{\i}}n}, {Tojeiro}, {Xu}, {Maga{\~n}a}, {Aubourg},
  {Bahcall}, {Bailey}, {Bizyaev}, {Bolton}, {Brewington}, {Brinkmann},
  {Brownstein}, {Gott}, {Hamilton}, {Ho}, {Honscheid}, {Labatie},
  {Malanushenko}, {Malanushenko}, {Maraston}, {Muna}, {Nichol}, {Oravetz},
  {Pan}, {Ross}, {Roe}, {Reid}, {Schlegel}, {Shelden}, {Schneider}, {Simmons},
  {Skibba}, {Snedden}, {Thomas}, {Tinker}, {Wake}, {Weaver}, {Weinberg},
  {White}, {Zehavi}, \& {Zhao}}]{Sanchez:2012}
{S{\'a}nchez}, A.~G., {Sc{\'o}ccola}, C.~G., {Ross}, A.~J., {Percival}, W.,
  {Manera}, M., {Montesano}, F., {Mazzalay}, X., {Cuesta}, A.~J., {Eisenstein},
  D.~J., {Kazin}, E., {McBride}, C.~K., {Mehta}, K., {Montero-Dorta}, A.~D.,
  {Padmanabhan}, N., {Prada}, F., {Rubi{\~n}o-Mart{\'{\i}}n}, J.~A., {Tojeiro},
  R., {Xu}, X., {Maga{\~n}a}, M.~V., {Aubourg}, E., {Bahcall}, N.~A., {Bailey},
  S., {Bizyaev}, D., {Bolton}, A.~S., {Brewington}, H., {Brinkmann}, J.,
  {Brownstein}, J.~R., {Gott}, J.~R., {Hamilton}, J.~C., {Ho}, S., {Honscheid},
  K., {Labatie}, A., {Malanushenko}, E., {Malanushenko}, V., {Maraston}, C.,
  {Muna}, D., {Nichol}, R.~C., {Oravetz}, D., {Pan}, K., {Ross}, N.~P., {Roe},
  N.~A., {Reid}, B.~A., {Schlegel}, D.~J., {Shelden}, A., {Schneider}, D.~P.,
  {Simmons}, A., {Skibba}, R., {Snedden}, S., {Thomas}, D., {Tinker}, J.,
  {Wake}, D.~A., {Weaver}, B.~A., {Weinberg}, D.~H., {White}, M., {Zehavi}, I.,
  \& {Zhao}, G. 2012, \mnras, 425, 415

\bibitem[{Santos {et~al.}(2015{\natexlab{a}})Santos, Silva, \&
  Yahya}]{SKA:HIsims}
Santos, M., Silva, M., \& Yahya, S. 2015{\natexlab{a}}, Proceedings of
  ``Advancing Astrophysics with the Square Kilometre Array'', PoS (AASKA15)

\bibitem[{Santos {et~al.}(2015{\natexlab{b}})}]{Santos:HIIM}
Santos, M. {et~al.} 2015{\natexlab{b}}, Proceedings of ``Advancing Astrophysics
  with the Square Kilometre Array'', PoS (AASKA15)

\bibitem[{Schmidt(2009)}]{Schmidt:2009}
Schmidt, F. 2009, Phys.Rev.D, 80, 123003

\bibitem[{Schmidt {et~al.}(2013)Schmidt, Jeong, \& Desjacques}]{Schmidt:2013}
Schmidt, F., Jeong, D., \& Desjacques, V. 2013, Physical Review D, 88, 23515

\bibitem[{Scoccimarro(2004)}]{Scoccimarro:2004}
Scoccimarro, R. 2004, astro-ph/0407214, astro-ph

\bibitem[{Seo \& Eisenstein(2007)}]{Seo:2006}
Seo, H.-J. \& Eisenstein, D.~J. 2007, Astrophys.J., 665, 14

\bibitem[{Simpson \& Peacock(2010)}]{Simpson:2010}
Simpson, F. \& Peacock, J.~A. 2010, Phys Rev D,, 81, 043512

\bibitem[{Smith {et~al.}(2003)Smith, Peacock, Jenkins, White, Frenk, Pearce,
  Thomas, Efstathiou, Couchmann, \& Consortium}]{Smith:2003}
Smith, R., Peacock, J., Jenkins, A., White, S., Frenk, C., Pearce, F., Thomas,
  P., Efstathiou, G., Couchmann, H., \& Consortium, V. 2003,
  Mon.Not.Roy.Astron.Soc., 341, 1311

\bibitem[{Smith {et~al.}(2006)Smith, Scoccimarro, \& Sheth}]{Smith:2006}
Smith, R.~E., Scoccimarro, R., \& Sheth, R.~K. 2006, astro-ph/0609547

\bibitem[{Song \& Koyama(2008)}]{Song:2008b}
Song, Y.-S. \& Koyama, K. 2008, arXiv:0802.3897, astro-ph

\bibitem[{Song \& Percival(2008)}]{Song:2008a}
Song, Y.-S. \& Percival, W.~J. 2008, arXiv:0807.0810, astro-ph

\bibitem[{Szalay {et~al.}(1997)Szalay, Matsubara, \& Landy}]{Szalay:1997}
Szalay, A.~S., Matsubara, T., \& Landy, S.~D. 1997, astro-ph/9712007, astro-ph

\bibitem[{Szapudi(2004)}]{Szapudi:2004}
Szapudi, I. 2004, The Astrophysical Journal, 614, 51

\bibitem[{Taruya {et~al.}(2010)Taruya, Nishimichi, \& Saito}]{Taruya:2010}
Taruya, A., Nishimichi, T., \& Saito, S. 2010, arXiv:1006.0699

\bibitem[{Taruya {et~al.}(2009)Taruya, Nishimichi, Saito, \&
  Hiramatsu}]{Taruya:2009}
Taruya, A., Nishimichi, T., Saito, S., \& Hiramatsu, T. 2009, arXiv:0906.0507

\bibitem[{{Tegmark} {et~al.}(2006){Tegmark}, {Eisenstein}, {Strauss},
  {Weinberg}, {Blanton}, {Frieman}, {Fukugita}, {Gunn}, {Hamilton}, {Knapp},
  {Nichol}, {Ostriker}, {Padmanabhan}, {Percival}, {Schlegel}, {Schneider},
  {Scoccimarro}, {Seljak}, {Seo}, {Swanson}, {Szalay}, {Vogeley}, {Yoo},
  {Zehavi}, {Abazajian}, {Anderson}, {Annis}, {Bahcall}, {Bassett}, {Berlind},
  {Brinkmann}, {Budavari}, {Castander}, {Connolly}, {Csabai}, {Doi},
  {Finkbeiner}, {Gillespie}, {Glazebrook}, {Hennessy}, {Hogg}, {Ivezi{\'c}},
  {Jain}, {Johnston}, {Kent}, {Lamb}, {Lee}, {Lin}, {Loveday}, {Lupton},
  {Munn}, {Pan}, { Park}, {Peoples}, {Pier}, {Pope}, {Richmond}, {Rockosi},
  {Scranton}, {Sheth}, {Stebbins}, {Stoughton}, {Szapudi}, {Tucker}, {vanden
  Berk}, {Yanny}, \& {York}}]{Tegmark:2006}
{Tegmark}, M., {Eisenstein}, D.~J., {Strauss}, M.~A., {Weinberg}, D.~H.,
  {Blanton}, M.~R., {Frieman}, J.~A., {Fukugita}, M., {Gunn}, J.~E.,
  {Hamilton}, A.~J.~S., {Knapp}, G.~R., {Nichol}, R.~C., {Ostriker}, J.~P.,
  {Padmanabhan}, N., {Percival}, W.~J., {Schlegel}, D.~J., {Schneider}, D.~P.,
  {Scoccimarro}, R., {Seljak}, U., {Seo}, H., {Swanson}, M., {Szalay}, A.~S.,
  {Vogeley}, M.~S., {Yoo}, J., {Zehavi}, I., {Abazajian}, K., {Anderson},
  S.~F., {Annis}, J., {Bahcall}, N.~A., {Bassett}, B., {Berlind}, A.,
  {Brinkmann}, J., {Budavari}, T., {Castander}, F., {Connolly}, A., {Csabai},
  I., {Doi}, M., {Finkbeiner}, D.~P., {Gillespie}, B., {Glazebrook}, K.,
  {Hennessy}, G.~S., {Hogg}, D.~W., {Ivezi{\'c}}, {\v Z}., {Jain}, B.,
  {Johnston}, D., {Kent}, S., {Lamb}, D.~Q., {Lee}, B.~C., {Lin}, H.,
  {Loveday}, J., {Lupton}, R.~H., {Munn}, J.~A., {Pan}, K., { Park}, C.,
  {Peoples}, J., {Pier}, J.~R., {Pope}, A., {Richmond}, M., {Rockosi}, C.,
  {Scranton}, R., {Sheth}, R.~K., {Stebbins}, A., {Stoughton}, C., {Szapudi},
  I., {Tucker}, D.~L., {vanden Berk}, D.~E., {Yanny}, B., \& {York}, D.~G.
  2006, Phys.Rev.D, 74, 123507

\bibitem[{Tegmark {et~al.}(1997)Tegmark, Taylor, \& Heavens}]{Tegmark:1997}
Tegmark, M., Taylor, A., \& Heavens, A. 1997, Astrophys.J., 480, 22

\bibitem[{Xia {et~al.}(2010)Xia, Viel, Baccigalupi, Zotti, Matarrese, \&
  Verde}]{Xia:2010}
Xia, J.-Q., Viel, M., Baccigalupi, C., Zotti, G.~D., Matarrese, S., \& Verde,
  L. 2010, Astrophys. J., 717, L17

\bibitem[{Yamamoto {et~al.}(2010)Yamamoto, Nakamura, Huetsi, Narikawa, \&
  Sato}]{Yamamoto:2010}
Yamamoto, K., Nakamura, G., Huetsi, G., Narikawa, T., \& Sato, T. 2010,
  Phys.Rev.D, 81, 103517

\bibitem[{Yamamoto {et~al.}(2008)Yamamoto, Sato, \& Huetsi}]{Yamamoto:2008}
Yamamoto, K., Sato, T., \& Huetsi, G. 2008, Prog.Theor.Phys., 120, 609

\bibitem[{Yoo(2010)}]{Yoo:2010}
Yoo, J. 2010, Phys.Rev.D, 82, 083508

\bibitem[{Yoo {et~al.}(2012)Yoo, Hamaus, Seljak, \& Zaldarriaga}]{Yoo:2012}
Yoo, J., Hamaus, N., Seljak, U., \& Zaldarriaga, M. 2012, Phys.Rev.D, 86,
  063514

\bibitem[{Zhao {et~al.}(2012)Zhao, Saito, Percival, Ross, Montesano, Viel,
  Schneider, Ernst, Manera, Miralda-Escude, Ross, Samushia, Sanchez, Swanson,
  Thomas, Tojeiro, Yeche, \& York}]{Zhao:2012}
Zhao, G.-B., Saito, S., Percival, W.~J., Ross, A.~J., Montesano, F., Viel, M.,
  Schneider, D.~P., Ernst, D.~J., Manera, M., Miralda-Escude, J., Ross, N.~P.,
  Samushia, L., Sanchez, A.~G., Swanson, M. E.~C., Thomas, D., Tojeiro, R.,
  Yeche, C., \& York, D.~G. 2012, arXiv:1211.3741

\end{thebibliography}

\end{document}